\documentclass[conference]{IEEEtran}
\usepackage{cite}
\usepackage{amsmath,amssymb}
\usepackage{booktabs}
\usepackage{array}
\usepackage{graphicx}
\usepackage{multirow}
\usepackage{url}
\usepackage[hidelinks]{hyperref}

\graphicspath{{figures/}}

\title{Image-Based Malware Type Classification on MalNet-Image Tiny: Effects of Multi-Scale Fusion, Transfer Learning, Data Augmentation, and Schedule-Free Optimization}

\author{%
\IEEEauthorblockN{Ahmed A. Abouelkhaire, Waleed A. Yousef, and Issa Traor\'e}
\IEEEauthorblockA{Department of Electrical and Computer Engineering\\
University of Victoria\\
Victoria, BC, Canada}
}

\begin{document}
\maketitle

\begin{abstract}
This paper studies 43-class malware type classification on MalNet-Image Tiny, a public benchmark derived from Android APK files. The goal is to assess whether a compact image classifier benefits from four components evaluated in a controlled ablation: a feature pyramid network (FPN) for scale variation induced by resizing binaries of different lengths, ImageNet pretraining, lightweight augmentation through Mixup and TrivialAugment, and schedule-free AdamW optimization. All experiments use a ResNet18 backbone and the provided train/validation/test split. Reproducing the benchmark-style configuration yields macro-F1, denoted $F1_{\mathrm{macro}}$, of $0.6510$, consistent with the reported baseline of approximately $0.65$. Replacing the optimizer with schedule-free AdamW and using unweighted cross-entropy increases $F1_{\mathrm{macro}}$ to $0.6535$ in 10 epochs, compared with 96 epochs for the reproduced baseline. The best configuration combines pretraining, Mixup, TrivialAugment, and FPN, reaching $F1_{\mathrm{macro}}=0.6927$, $P_{\mathrm{macro}}=0.7707$, $\mathrm{AUC}_{\mathrm{macro}}=0.9556$, and $\mathcal{L}_{\mathrm{test}}=0.8536$. The ablation indicates that the largest gains in $F1_{\mathrm{macro}}$ arise from pretraining and augmentation, whereas FPN mainly improves $P_{\mathrm{macro}}$, $\mathrm{AUC}_{\mathrm{macro}}$, and $\mathcal{L}_{\mathrm{test}}$ in the strongest configuration.
\end{abstract}

\begin{IEEEkeywords}
malware classification, static analysis, malware images, MalNet-Image Tiny, transfer learning, feature pyramid network, AdamW schedule-free
\end{IEEEkeywords}

\section{Introduction}
Malware classification remains a difficult supervised learning problem because available corpora are heterogeneous, class distributions are imbalanced, and malicious software evolves faster than manually engineered feature sets. Image-based static analysis offers a simple alternative representation pipeline: binaries are converted into images and processed by vision models without code execution. This line of work traces back to malware visualization, which showed that samples from the same family often exhibit similar textures and structural patterns in image form~\cite{nataraj_malware_2011}. More recently, MalNet-Image introduced a large public benchmark with more than 1.2 million malware images organized into 696 families and 47 types, together with the smaller MalNet-Image Tiny subset for faster experimentation~\cite{freitas_malnet_2022}.

This paper evaluates four components that have not been jointly studied on MalNet-Image Tiny in the cited literature: i) a feature pyramid network (FPN) to mitigate scale variation introduced when binaries of different lengths are resized to a common image size, ii) schedule-free AdamW to reduce reliance on an explicit learning-rate schedule, iii) ImageNet pretraining, and iv) lightweight augmentation through Mixup and TrivialAugment. The full 47-type benchmark remains difficult: the MalNet-Image study reported type-level macro-F1 values of only $0.47$--$0.48$ for standard CNN backbones despite substantial differences in model size~\cite{freitas_malnet_2022}. The objective here is therefore narrower and more controlled: to quantify the contribution of training strategy and multi-scale fusion on the fixed MalNet-Image Tiny split while keeping the backbone architecture compact.

\section{Related Work and Dataset Context}

Prior work falls into three main groups.

\subsection{Sequence- and language-based malware modeling}
A first line of work treats opcode or API-call sequences as text-like data. Demirkiran \emph{et al.} used ensembles of pre-trained transformer models, including BERT and CANINE, and reported strong multiclass performance on Catak, Oliveira, VirusSample, and VirusShare datasets~\cite{demirkiran_ensemble_2022}. Yesir and colleagues compared fastText and BERT on API-call sequences extracted by dynamic analysis and found that fastText was often more efficient while remaining competitive on CSDMC, APIMDS, and a custom dataset~\cite{yesir_malware_2021}. Zhang \emph{et al.} proposed CoDroid, which combines static opcode sequences and dynamic system-call sequences through a CNN--BiLSTM--attention architecture for Android malware detection~\cite{zhang_hybrid_2021}. These methods can capture sequential semantics, but they depend on sequence extraction pipelines and are not directly comparable to raw-image classification.

\subsection{Graph-based malware modeling}
A second line of work represents programs or behaviors as graphs. DLGraph combines function-call graphs and Windows API calls by embedding graphs with node2vec and then learning joint representations with stacked denoising autoencoders~\cite{jiang_dlgraph_2018}. AMalNet integrates graph convolutional networks with Independently Recurrent Neural Networks to model API-call graphs and remote dependencies at scale~\cite{pei_amalnet_2020}. GDroid formulates Android malware detection and family attribution as node classification in a heterogeneous graph connecting applications and Android APIs~\cite{gao_gdroid_2021}. Graph methods capture structural dependencies that image models do not represent explicitly, but they also require graph construction and domain-specific parsers.

\subsection{Image-based malware classification}
Nataraj \emph{et al.} introduced malware visualization and grayscale image classification using GIST descriptors on 9{,}458 samples from 25 families~\cite{nataraj_malware_2011}. Alguliyev \emph{et al.} combined grayscale malware images, the Radon transform, and transfer learning, reporting high accuracy on Microsoft Malware, IoT-Malware, and MalNet-Image subsets~\cite{alguliyev_radon_2024}. STAMINA addressed file-size heterogeneity through file-size gating and separate image-width rules, then trained deep models on the resulting malware-as-image representation~\cite{chen2020stamina}. Kumar \emph{et al.} evaluated fine-tuned ImageNet backbones such as VGG16, VGG19, ResNet50, and InceptionV3 for malware image classification~\cite{kumar_sdif-cnn_2023}. Three gaps motivate this paper: limited use of the standard MalNet-Image Tiny split for controlled ablation, limited study of lightweight data-independent augmentation in malware imagery, and limited investigation of end-to-end multi-scale handling without manually deploying multiple models.

\subsection{Dataset coverage}
Table~\ref{tab:datasets} summarizes image-based malware datasets relevant to this work. MalNet-Image Tiny is used because it preserves the public benchmark pipeline, including the fixed split protocol, while excluding the four largest types from the full dataset. This reduces computational cost and moderates, but does not remove, the class imbalance inherited from MalNet-Image.

\begin{table}[t]
\caption{Image-based malware datasets relevant to this work.}
\label{tab:datasets}
\centering
\footnotesize
\setlength{\tabcolsep}{3pt}
\begin{tabular}{lcrr}
\toprule
Dataset & Access & Images & Classes \\
\midrule
MalNet-Image~\cite{freitas_malnet_2022} & Public & 1{,}262{,}024 & 696 \\
MalNet-Image Tiny~\cite{freitas_malnet_2022} & Public & 87{,}430 & 43 \\
Virus-MNIST~\cite{noever2021virusmnist} & Public & 51{,}880 & 10 \\
Malimg~\cite{nataraj_malware_2011} & Public & 9{,}458 & 25 \\
AndroDex~\cite{aurangzeb2024androdex} & Public & 24{,}746 & 179 \\
STAMINA~\cite{chen2020stamina} & Private & 782{,}224 & 2 \\
McAfee~\cite{mcafee2020} & Private & 367{,}183 & 2 \\
\bottomrule
\end{tabular}
\end{table}

\section{Materials and Methods}

\subsection{Dataset and conversion pipeline}
MalNet-Image was constructed from Android APK files obtained from AndroZoo and labeled with Euphony, which aggregates antivirus labels from VirusTotal vendors~\cite{freitas_malnet_2022}. The experiments use MalNet-Image Tiny, which contains 87{,}430 images from 43 classes with fixed splits of 61{,}201 for training, 8{,}743 for validation, and 17{,}486 for testing; these counts follow the benchmark's 70/10/20 split design. The Tiny subset excludes the four largest types from the full benchmark and is therefore less imbalanced, although imbalance remains evident. The thesis source also notes apparent label inconsistencies, including near-duplicate type names such as ``adsware'' and ``adware,'' which may introduce additional label noise.

\begin{figure}[t]
\centering
\includegraphics[width=\columnwidth]{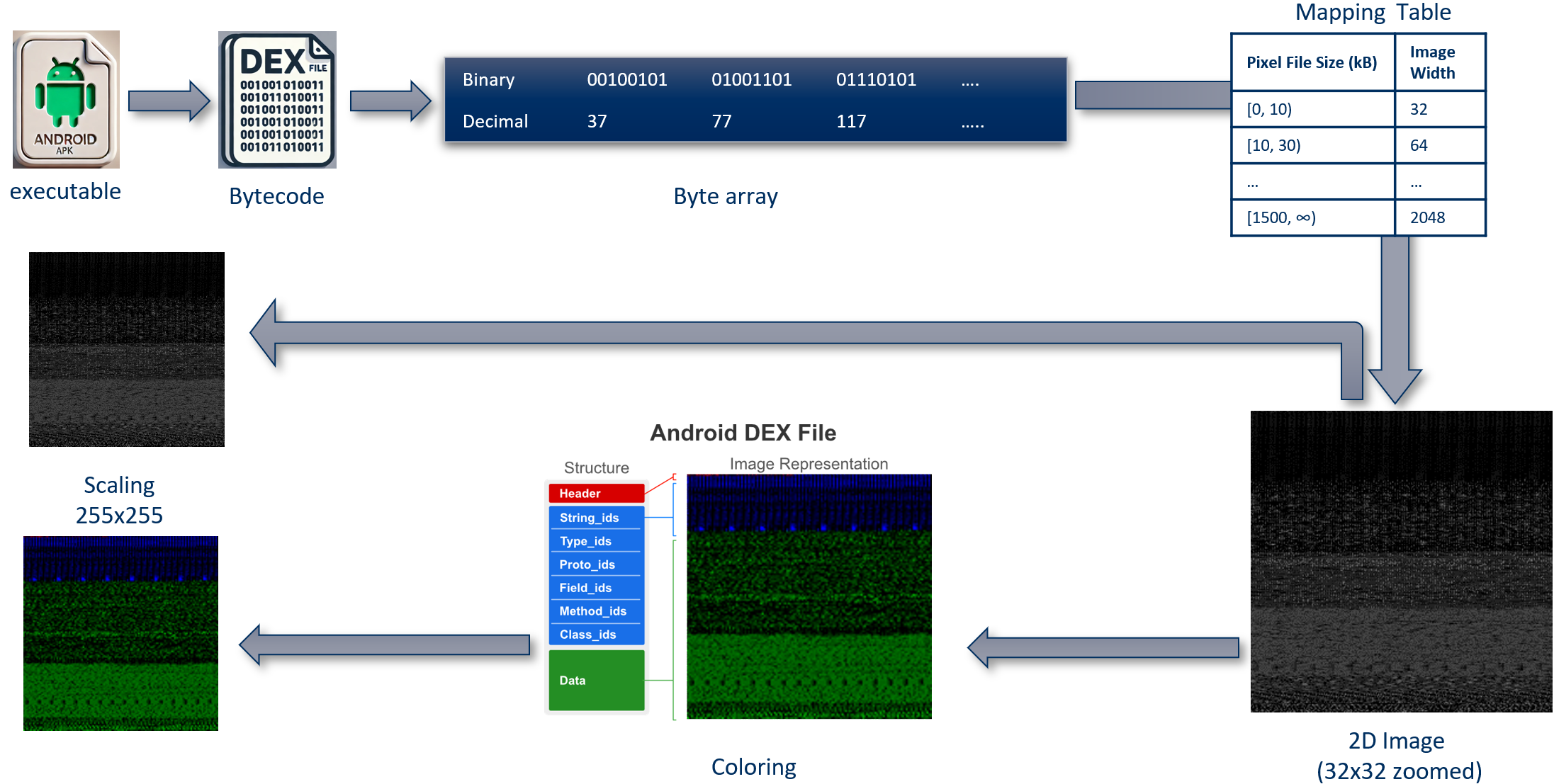}
\caption{APK-to-image conversion pipeline. The DEX byte stream is reshaped into a two-dimensional grid and resized to $256\times256$; the three-channel variant assigns colors by DEX section.}
\label{fig:conversion}
\end{figure}

Figure~\ref{fig:conversion} summarizes the image-conversion pipeline. The DEX file is extracted from the APK, converted into a one-dimensional array of 8-bit unsigned integers, reshaped into a two-dimensional grid according to the file-to-pixel-width mapping used in the dataset construction process, and then resized to $256\times256$ with a Lanczos filter. Two input variants are evaluated. The first uses a single grayscale channel ($K=1$). The second uses three channels ($K=3$) and color-codes byte positions by DEX section, including the header, identifier, class-definition, and data regions.

\subsection{Backbone, multi-scale extension, and transfer learning}
ResNet18 is used as the backbone because the MalNet benchmark reported a favorable trade-off between type-level performance and computational cost for this architecture~\cite{freitas_malnet_2022,he_deep_2016}. On the full benchmark, ResNet18 reached type-level macro-F1 of $0.47$ with 12M parameters and 1.8 GFLOPs, whereas ResNet50 and ResNet101 reached $0.48$ with 26M and 45M parameters and 3.9 and 7.6 GFLOPs, respectively~\cite{freitas_malnet_2022}. These results motivate the use of ResNet18 as a compact backbone for isolating the effects of optimization, augmentation, and multi-scale fusion.

To address scale changes induced by resizing binaries of different lengths, some experiments attach an FPN to intermediate ResNet stages. Let $\{C_2,C_3,C_4,C_5\}$ denote the selected backbone feature maps and $\{P_2,P_3,P_4,P_5\}$ the corresponding pyramid outputs, with $P_5=\mathrm{Conv}_{1\times1}(C_5)$. For $i\in\{2,3,4\}$, the top-down fusion rule is
\begin{equation}
P_i = \mathrm{Conv}_{1\times1}(C_i) + \mathrm{Upsample}(P_{i+1}),
\end{equation}
where $\mathrm{Conv}_{1\times1}(\cdot)$ denotes a lateral $1\times1$ projection and $\mathrm{Upsample}(\cdot)$ doubles the spatial resolution. ImageNet pretraining is evaluated because prior work reported that pretrained visual representations remain useful even when the target images differ substantially from natural photographs~\cite{kumar_sdif-cnn_2023,yosinski_2014_transferable}.

\subsection{Optimization, losses, and augmentation}
The reproduced benchmark baseline uses AdamW, weighted cross-entropy, grayscale input, and ResNet18~\cite{freitas_malnet_2022}. The main optimizer studied here is schedule-free AdamW~\cite{defazio_road_2024}. Let $g_t$ denote the stochastic gradient at iteration $t$, and let $x_t$, $y_t$, and $z_t$ denote the parameter sequences used by the optimizer. The second-moment estimate is
\begin{equation}
v_t = \beta_2 v_{t-1} + (1-\beta_2) g_t^2, \qquad \hat v_t = \frac{v_t}{1-\beta_2^t},
\end{equation}
and the schedule-free update is
\begin{align}
y_t &= (1-\beta_1) z_t + \beta_1 x_t, \\
z_{t+1} &= z_t - \frac{\eta_t}{\sqrt{\hat v_t}+\epsilon} g_t - \eta_t\lambda_{\mathrm{wd}} y_t, \\
x_{t+1} &= (1-c_{t+1})x_t + c_{t+1}z_{t+1},
\end{align}
with
\begin{equation}
c_{t+1}=\frac{\eta_t^2}{\sum_{i=1}^{t}\eta_i^2}, \qquad
\eta_t=\eta\min\left(1,\frac{t}{T_{\mathrm{warmup}}}\right).
\end{equation}
This optimizer is studied because it removes an explicit learning-rate schedule and the original method reports sensitivity to $\beta_1$, which was therefore included in the hyperparameter search.

For class imbalance, both standard cross-entropy and weighted cross-entropy were evaluated. For a batch of size $B$ and $C$ classes, let $p_{bc}$ denote the predicted probability of class $c$ for example $b$, and let $y_{bc}$ denote the corresponding one-hot target. The weighted loss is
\begin{equation}
\mathcal{L}_{\mathrm{WCE}}=-\frac{1}{B}\sum_{b=1}^{B}\sum_{c=1}^{C} w_c\, y_{bc}\log p_{bc}, \qquad
w_c=\frac{N}{C n_c},
\end{equation}
where $N$ is the total number of training examples and $n_c$ is the number of training examples in class $c$. Standard cross-entropy is recovered by setting $w_c=1$ for all classes.

The augmentation methods are Mixup~\cite{zhang_mixup_2018}, which forms convex combinations of two training pairs $(\mathbf{x}^{(a)},\mathbf{y}^{(a)})$ and $(\mathbf{x}^{(b)},\mathbf{y}^{(b)})$,
\begin{align}
\tilde{\mathbf{x}} &= \lambda_{\mathrm{mix}}\mathbf{x}^{(a)} + (1-\lambda_{\mathrm{mix}})\mathbf{x}^{(b)}, \\
\tilde{\mathbf{y}} &= \lambda_{\mathrm{mix}}\mathbf{y}^{(a)} + (1-\lambda_{\mathrm{mix}})\mathbf{y}^{(b)}.
\end{align}
where $\lambda_{\mathrm{mix}}\in[0,1]$, and TrivialAugment~\cite{mueller_trivialaugment_2021}, which applies a single randomly selected augmentation operation with a sampled magnitude and does not require a search procedure.

\subsection{Evaluation protocol}
The experiments follow the benchmark-style fixed split. Performance is reported using macro-precision $P_{\mathrm{macro}}$, macro-recall $R_{\mathrm{macro}}$, macro-F1 $F1_{\mathrm{macro}}$, and macro-AUC $\mathrm{AUC}_{\mathrm{macro}}$ because macro-averaging assigns equal weight to each class and is therefore more appropriate than accuracy under class imbalance. For any class-wise metric $M_c$ over $C=43$ classes, the macro average is
\begin{equation}
M_{\mathrm{macro}}=\frac{1}{C}\sum_{c=1}^{C} M_c.
\end{equation}
This convention is used consistently for precision, recall, F1, and AUC. The selected checkpoint for each configuration is determined by validation $F1_{\mathrm{macro}}$, with validation loss used only to break ties. Both optimization loss and $F1_{\mathrm{macro}}$ are reported because lower loss does not necessarily imply better class-balanced performance.

\begin{figure}[t]
\centering
\includegraphics[width=\columnwidth]{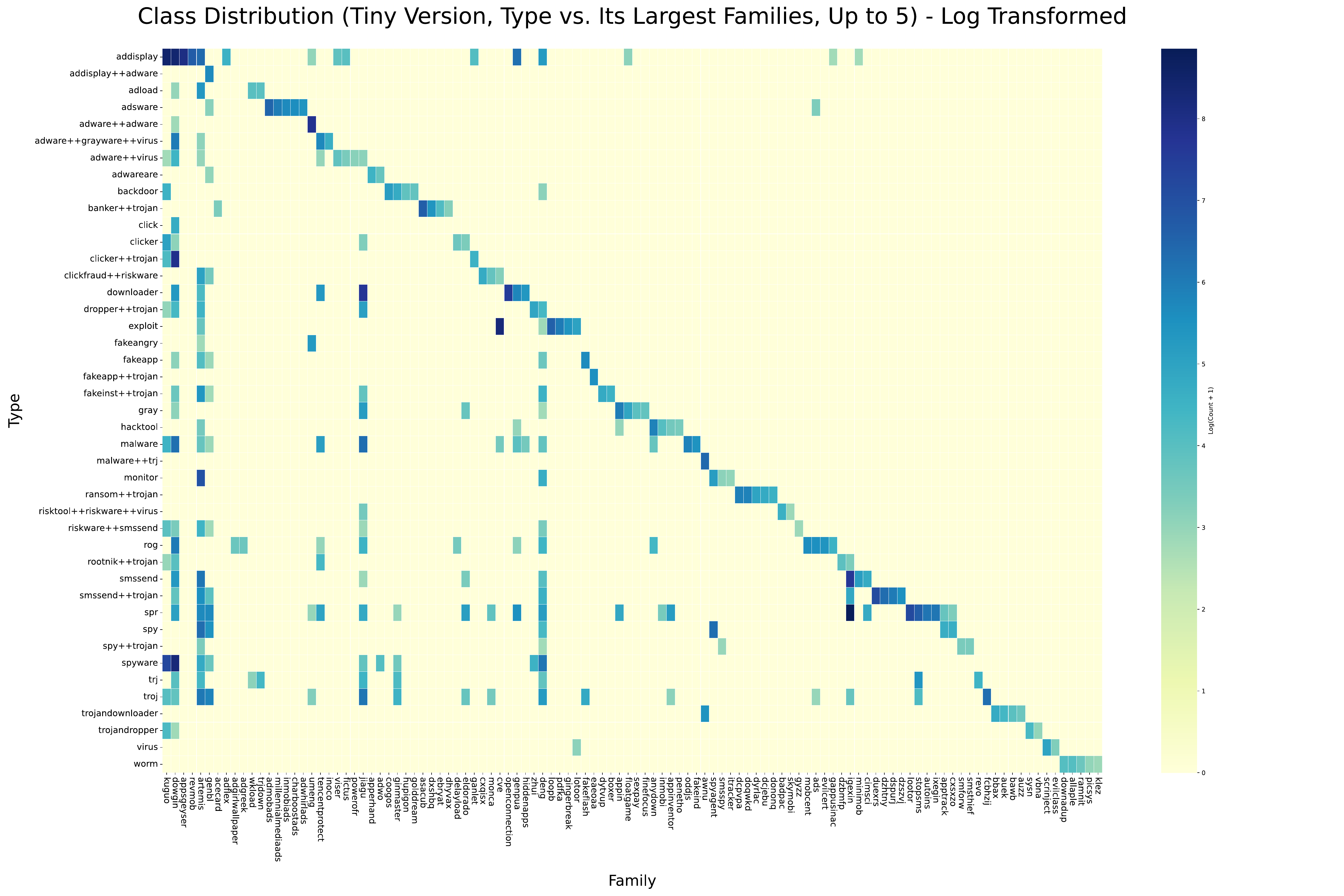}
\caption{MalNet-Image Tiny class distribution. The Tiny subset is less imbalanced than the full MalNet-Image benchmark, but substantial imbalance remains across type--family pairs.}
\label{fig:imbalance}
\end{figure}

\section{Results}

\subsection{Experimental design and hyperparameter search}
The 15 experiments form a staged ablation. The sequence is: reproduce the published baseline; replace AdamW with schedule-free AdamW; compare weighted and unweighted cross-entropy; assess ImageNet pretraining; assess Mixup and TrivialAugment on grayscale and three-channel inputs; assess FPN; and finally combine the strongest settings. The schedule-free hyperparameter search covers learning rates $\{0.01, 0.001, 0.005\}$, weight decay $0.01$, warmup steps $1000$, $\beta_1\in\{0.9,0.95\}$, $\beta_2=0.999$, and batch size $128$. The stronger schedule-free runs used $\beta_1=0.95$.

\begin{table}[t]
\caption{Hyperparameter values explored for schedule-free AdamW.}
\label{tab:hyper}
\centering
\footnotesize
\begin{tabular}{ll}
\toprule
Hyperparameter & Values \\
\midrule
Learning rate & 0.01, 0.001, 0.005 \\
Weight decay & 0.01 \\
Warmup steps & 1000 \\
$\beta_1$ & 0.9, 0.95 \\
$\beta_2$ & 0.999 \\
Batch size & 128 \\
\bottomrule
\end{tabular}
\end{table}

\subsection{Full ablation results}
Table~\ref{tab:ablationfull} reports the complete experiment set. Experiment~5 reproduces the benchmark-style baseline with $F1_{\mathrm{macro}}=0.6510$, which is consistent with the value reported for ResNet18 on MalNet-Image Tiny~\cite{freitas_malnet_2022}. Experiment~6 shows that schedule-free AdamW with unweighted cross-entropy slightly exceeds this result at $F1_{\mathrm{macro}}=0.6535$ while requiring 10 epochs rather than 96. Across Experiments~9--15, the largest incremental improvements are associated with pretraining and augmentation. The strongest non-FPN model is Experiment~14 with $F1_{\mathrm{macro}}=0.6927$, and the strongest overall configuration is Experiment~15, which preserves the same $F1_{\mathrm{macro}}$ while improving $P_{\mathrm{macro}}$, $\mathrm{AUC}_{\mathrm{macro}}$, and $\mathcal{L}_{\mathrm{test}}$.

\begin{table*}[t]

\caption{Complete ablation table. PT: pretrained; In: input channels ($1$ = grayscale, $3$ = RGB-like DEX color coding); TA: TrivialAugment; MU: Mixup; AF: schedule-free AdamW; AW: AdamW; CE: cross-entropy; WCE: weighted cross-entropy. Metrics are $P_{\mathrm{macro}}$, $R_{\mathrm{macro}}$, $F1_{\mathrm{macro}}$, and $\mathrm{AUC}_{\mathrm{macro}}$.}
\label{tab:ablationfull}
\centering
\scriptsize
\setlength{\tabcolsep}{3.2pt}
\begin{tabular}{ccccccccccccc}
\toprule
ID & PT & FPN & In & TA & MU & Opt & Loss & $P_{\mathrm{macro}}$ & $R_{\mathrm{macro}}$ & $F1_{\mathrm{macro}}$ & $\mathrm{AUC}_{\mathrm{macro}}$ & $\mathcal{L}_{\mathrm{test}}$ \\
\midrule
1  & N & N & 1 & N & N & AF & WCE & 0.6455 & 0.6266 & 0.6276 & 0.9487 & 1.7064 \\
2  & Y & N & 3 & N & N & AW & CE  & 0.6978 & 0.5975 & 0.6315 & 0.9434 & 1.2115 \\
3  & N & N & 1 & N & N & AF & CE  & 0.6715 & 0.6083 & 0.6317 & 0.9409 & 1.7247 \\
4  & N & N & 1 & N & N & AF & CE  & 0.6743 & 0.6320 & 0.6449 & 0.9475 & 1.1375 \\
5  & N & N & 1 & N & N & AW & WCE & 0.6651 & 0.6460 & 0.6510 & 0.9488 & 1.8410 \\
6  & N & N & 1 & N & N & AF & CE  & 0.6928 & 0.6301 & 0.6535 & 0.9451 & 1.5553 \\
7  & Y & N & 3 & N & N & AW & CE  & 0.6907 & 0.6390 & 0.6536 & 0.9512 & 1.3389 \\
8  & Y & Y & 3 & N & N & AF & CE  & 0.6936 & 0.6499 & 0.6654 & 0.9492 & 1.3520 \\
9  & Y & N & 1 & N & N & AF & CE  & 0.6951 & 0.6507 & 0.6656 & 0.9503 & 1.1690 \\
10 & Y & N & 3 & N & Y & AF & CE  & 0.7276 & 0.6411 & 0.6698 & 0.9519 & 0.8586 \\
11 & Y & N & 1 & N & Y & AF & CE  & 0.7315 & 0.6424 & 0.6728 & 0.9514 & 0.8618 \\
12 & N & Y & 3 & Y & Y & AF & CE  & 0.7694 & 0.6269 & 0.6755 & 0.9521 & 0.9023 \\
13 & Y & N & 3 & Y & N & AF & CE  & 0.7098 & 0.6582 & 0.6764 & 0.9515 & 1.0704 \\
14 & Y & N & 3 & Y & Y & AF & CE  & 0.7670 & 0.6537 & 0.6927 & 0.9552 & 0.8731 \\
15 & Y & Y & 3 & Y & Y & AF & CE  & 0.7707 & 0.6523 & 0.6927 & 0.9556 & 0.8536 \\
\bottomrule
\end{tabular}
\end{table*}

\subsection{Observed effects of each component}
The reproduced baseline indicates that the implementation is consistent with the benchmark protocol. Comparing Experiments~1, 3, 4, and 6 suggests that weighted cross-entropy does not provide a clear advantage under the schedule-free optimizer on the Tiny subset, which is less imbalanced than the full benchmark. Comparing Experiments~6, 9, and 13--15 indicates that pretrained weights are beneficial and that augmentation produces the largest gains in $F1_{\mathrm{macro}}$. Relative to Experiment~9, Mixup alone increases $F1_{\mathrm{macro}}$ to $0.6698$--$0.6728$, depending on the input representation; TrivialAugment alone increases $F1_{\mathrm{macro}}$ to $0.6764$; and the combination reaches $0.6927$. Adding FPN to the strongest pretrained and augmented configuration preserves $F1_{\mathrm{macro}}$ but improves $P_{\mathrm{macro}}$ from $0.7670$ to $0.7707$, $\mathrm{AUC}_{\mathrm{macro}}$ from $0.9552$ to $0.9556$, and $\mathcal{L}_{\mathrm{test}}$ from $0.8731$ to $0.8536$.

\begin{figure}[t]

\centering
\includegraphics[width=\columnwidth]{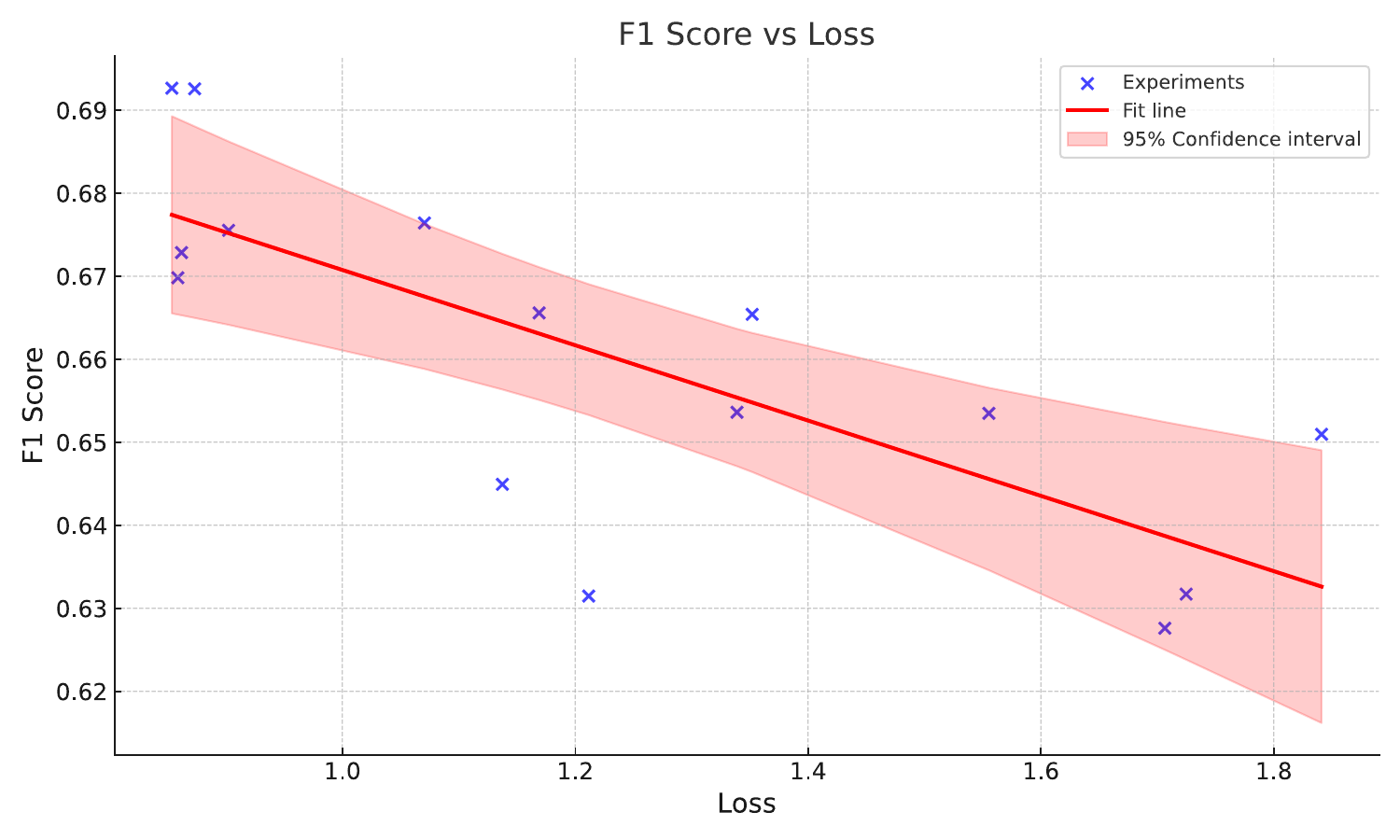}
\caption{$F1_{\mathrm{macro}}$ versus $\mathcal{L}_{\mathrm{test}}$ across the ablation runs. Lower loss is associated with, but does not determine, higher $F1_{\mathrm{macro}}$.}
\label{fig:f1loss}
\end{figure}

\section{Discussion}
The evidence in Table~\ref{tab:ablationfull} suggests that the observed improvement is cumulative rather than attributable to a single change. Schedule-free AdamW mainly reduces the number of epochs required to reach baseline-level $F1_{\mathrm{macro}}$. The larger gains are associated with ImageNet initialization and lightweight augmentation, which is consistent with prior reports that transfer learning can be effective for malware-image classification~\cite{kumar_sdif-cnn_2023,alguliyev_radon_2024}. The comparison between grayscale and three-channel inputs is also informative: both representations remain competitive, and the results do not indicate a uniform advantage for one encoding across all settings. Instead, the strongest results arise when the input representation is combined with pretraining and augmentation rather than evaluated in isolation.

The FPN hypothesis is only partially supported. Fixed resizing can change the apparent scale of similar bytecode regions across binaries of different lengths. Empirically, FPN improves $P_{\mathrm{macro}}$, $\mathrm{AUC}_{\mathrm{macro}}$, and $\mathcal{L}_{\mathrm{test}}$ under the strongest configuration, but it does not increase $F1_{\mathrm{macro}}$ beyond the best non-FPN model. This pattern suggests that multi-scale fusion may improve score calibration and class separation without materially changing the class-balanced error profile summarized by $F1_{\mathrm{macro}}$. It is also consistent with the decision to study training strategy on top of a compact backbone rather than scaling depth alone, because the benchmark comparison among standard ResNet variants shows only small type-level differences at substantially higher computational cost~\cite{freitas_malnet_2022}.

Several limitations remain important. First, the experiments are systematic but not exhaustive. Second, evaluation is restricted to the provided MalNet-Image Tiny split rather than repeated resampling or cross-validation. Third, the dataset contains apparent label inconsistencies, such as near-duplicate type names, that may limit achievable performance. Fourth, optimization remains indirect because $F1_{\mathrm{macro}}$ is not used as the training objective. Figure~\ref{fig:f1loss} illustrates that lower loss does not always imply higher $F1_{\mathrm{macro}}$. Finally, the reproduced performance on Tiny should not be interpreted as directly comparable to the full 47-type benchmark, because Tiny removes the four largest types and is less severely imbalanced.

\section{Conclusion}
On MalNet-Image Tiny, the reproduced ResNet18 baseline reaches $F1_{\mathrm{macro}}=0.6510$. Replacing AdamW with schedule-free AdamW and using unweighted cross-entropy increases $F1_{\mathrm{macro}}$ to $0.6535$ while substantially reducing the number of training epochs. The best configuration combines ImageNet pretraining, Mixup, TrivialAugment, and FPN, yielding $F1_{\mathrm{macro}}=0.6927$, $P_{\mathrm{macro}}=0.7707$, $\mathrm{AUC}_{\mathrm{macro}}=0.9556$, and $\mathcal{L}_{\mathrm{test}}=0.8536$. Within the reported ablations, the principal gains arise from pretraining and augmentation, whereas FPN improves secondary metrics rather than $F1_{\mathrm{macro}}$. Future work includes evaluation on the full MalNet-Image benchmark, improved handling of imbalance, refinement of the binary-to-image transformation, and dataset cleaning for label consistency.

\bibliographystyle{IEEEtran}
\bibliography{UvicThesis}

\end{document}